\documentclass{article}
\usepackage{spconf,amsmath,graphicx}
\usepackage{color}
\usepackage{xcolor} 
\usepackage{todonotes} 
\usepackage{multirow}
\usepackage{booktabs}
\usepackage{makecell}
\usepackage[acronym]{glossaries}

\DeclareMathOperator*{\argmin}{arg\,min}

\newcommand{\etal}{\emph{et al.} }

\newcommand{\fref}[1]{Figure \ref{#1}}
\newcommand{\eref}[1]{Eq. \eqref{#1}}

\newacronym{stft}{STFT}{short-time Fourier transform}
\newacronym{stftm}{STFTM}{short-time Fourier transform magnitude}
\newacronym{vqvae}{VQ-VAE}{vector-quantized variational autoencoder}
\newacronym{dnn}{DNN}{deep neural network}
\newacronym{se}{SE}{speech enhancement}
\newacronym{ssl}{SSL}{self-supervised learning}
\newacronym{ae}{AE}{autoencoder}
\newacronym{vq}{VQ}{vector quantization}
\newacronym{cnn}{CNN}{convolutional neural network}
\newacronym{relu}{ReLU}{rectified linear unit}
\newacronym{mse}{MSE}{mean-squared error}
\newacronym{sisdr}{SI-SDR}{scale-invariant signal-to-distortion ratio}
\newacronym{pesq}{PESQ}{perceptual evaluation of speech quality}
\newacronym{snr}{SNR}{signal-to-noise ratio}

\title{Incorporating Real-world Noisy Speech in Neural-network-based Speech Enhancement Systems}

\twoauthors
  {Yangyang Xia$^*$\thanks{$^*$Work done during internship at FRL Research}}
    {Carnegie Mellon University\\
Dept. of Electrical and Computer Engineering\\
     Pittsburgh, PA 15213, USA \\
     \texttt{raymondxia@cmu.edu}}
  {Buye Xu, Anurag Kumar}
    {Facebook Reality Labs Research \\
     Redmond, WA 98052, USA \\
     \texttt{\{xub, anuragkr\}@fb.com}}

\begin{document}
\maketitle
\begin{abstract}
Supervised speech enhancement relies on parallel databases of degraded speech signals and their clean reference signals during training. This setting prohibits the use of real-world degraded speech data that may better represent the scenarios where such systems are used. In this paper, we explore methods that enable supervised speech enhancement systems to train on real-world degraded speech data. Specifically, we propose a semi-supervised approach for speech enhancement in which we first train a modified vector-quantized variational autoencoder that solves a source separation task. We then use this trained autoencoder to further train an enhancement network using real-world noisy speech data by computing a triplet-based unsupervised loss function. Experiments show promising results for incorporating real-world data in training speech enhancement systems.
\end{abstract}


\section{Introduction}
Supervised single-channel speech enhancement has seen considerable improvement in the last few years, primarily due to the use of deep neural networks (DNNs) \cite{wang2014training}. Training an effective \acrfull{se} system requires parallel databases of simulated degraded speech signals and their reference signals as the learning objective is often a function of the clean speech signals.
The performance of \acrshort{se} systems trained on such artificially generated noisy speech inputs depends heavily on (a) the variety and amount of noise recordings available, and (b) if the simulated degradation is realistic. While these supervised \acrshort{se} systems have surpassed non data-driven approaches by a large margin \cite{luo2019conv}, concerns around their generalization capabilities remain. Enabling SE systems to learn from real-world noisy speech can ensure that the networks are trained on real acoustical conditions rather than synthetic ones. Moreover, these data are readily available and can be obtained with relative ease.
Lastly, such methods can also enable a system trained on simulated data to adapt to a new environment. 

The primary challenge in incorporating real-world noisy speech for training SE systems is the lack of corresponding clean speech signals as training targets. A few recently proposed methods seek for alternative reference signals. Mixture-invariant training (MixIT) \cite{wisdom2020unsupervised} attempted unsupervised speaker separation by forcing the network to separate mixture of mixtures. However, it can suffer from over-separation problem. Following MixIT, Noisy-target Training \cite{fujimura2021noisy} treats real-world noisy speech data as reference and mixes them with noise signals to generate ``more noisy'' signals for training the SE system.

Another possibility to relax supervision is through the prediction or generation of pseudo ground truth. Although it is tempting to calculate the loss through a no-reference speech quality prediction network \cite{9054204}, experiments have shown that \acrshort{dnn}s might over-optimize one perceptual metric without necessarily improving others \cite{martin2018deep,zhao2018perceptually}, let alone a prediction of them. Wang \etal used a pair of generative adversarial networks to map speech signals from noisy to clean \cite{wang2020self}. The trained generator is then used to generate a pseudo reference signal. A similar setup was also proposed and studied by Xiang and Bao \cite{Xiang2020} with multiple learning objectives. These studies were inspired by unpaired image-to-image translation 
through cycle-consistency constraints \cite{zhu2017unpaired}. However, in ~\cite{wang2020self} the cycle-consistency constraint did not enforce clean speech embeddings and degraded speech embeddings to share the same latent space by using multiple encoders.

Generation of pseudo reference signals can also be done through a latent representation. In particular, methods based on \acrfull{ssl} frameworks can be used. In this framework, a speech signal is typically transformed to a latent space by an autoencoder. Then, \acrshort{ssl} tasks are assigned in the latent space to establish correlations between a measure taken in this space and a physically meaningful measure taken in the signal domain. For example, the context encoder learns to generate content of a masked region in an image based on its surrounding pixels \cite{pathak2016context}. 

In this paper, we propose two unsupervised loss functions for speech enhancement enabled by \acrlong{ssl}. These unsupervised loss functions do not require the reference clean speech and allow us to incorporate real-world noisy speech in the training process. Our semi-supervised approach consists of two stages. The first supervised stage includes a novel modification to the \acrfull{vqvae} that solves a source separation task using a corpus of \emph{paired} data. In the second semi-supervised stage, the learned \acrshort{vqvae} is used to transform any given degraded speech signal to a pseudo noise ground truth and a pseudo speech ground truth, respectively. We then construct unsupervised losses based on a triplet formulation using these estimated ground truths. These losses are used to train an enhancement system along with the supervised losses from the paired data. Note that, the framework is designed in a semi-supervised setting with the assumption that some amount of \emph{paired} data and potentially (much) more \emph{unpaired} (real-world) data are available during training. The \emph{unpaired} data can be real-world noisy speech recordings for which corresponding clean references are not available. 

\textbf{Organization of this paper.}
In Section 2, we provide some necessary background on supervised \gls{se} and \acrshort{vqvae}. We then describe our method in Section 3. Experimental setups are described in Section 4 and results are discussed in Section 5. Section 6 concludes our paper. 

\section{Background}
\subsection{Supervised DNN-based speech enhancement}
We assume that the observed degraded speech contains clean speech corrupted by additive noise. This relationship can be established in the \acrfull{stft} domain as
\begin{equation}
    X[t, k] = S[t, k] + N[t, k]
\end{equation}
where $X[t,k]$, $S[t,k]$, and $N[t,k]$ represent the \acrshort{stft} at frame $t$ and frequency index $k$ of the degraded speech, clean speech, and noise, respectively.
One common \acrshort{se} method is to train a \acrshort{dnn} to predict a magnitude gain $G[t,k]$, so that the \acrfull{stftm} of enhanced speech signal can be obtained by
\begin{equation}
    \left|\hat{S}[t,k]\right| = G[t, k]\left|X[t, k]\right|.
\end{equation}
Finally, the phase of the degraded signal is combined with the enhanced \acrshort{stftm} to reconstruct the enhanced speech signal through inverse \acrshort{stft}.

Conventionally, the paired sets $(X[t,k], S[t,k])$ are required during training. The supervised training involves a reconstruction loss,
\begin{equation}\label{eq:supervised_se}
    L_{s}(\widehat{\Vec{S}}, \Vec{S}) = d(\widehat{\Vec{S}}, \Vec{S}),
\end{equation}
where $\Vec{S}$ and $\widehat{\Vec{S}}$ denote the clean and enhanced \acrshort{stftm} in vector form, and $d(\cdot)$ is a distance measure such as the \acrfull{mse}.

\subsection{Encoder-Decoder in self-supervised learning}
Self-supervised learning (SSL) methods usually construct tasks in a learned representation space. These tasks can be solved without requiring any labels for a given dataset. The assumption usually is that the representation learned by solving these pretext tasks will be useful for the downstream tasks. We follow the well-known encoder-decoder framework to learn such representations from speech signals.
This autoencoding process can be described by
\begin{gather}
    \Vec{e} = \text{Encoder}(\Vec{f})\\
    \widehat{\Vec{f}} = \text{Decoder}(\Vec{e}) \\
    L_{\text{rec}}(\Vec{f}, \widehat{\Vec{f}}) = d(\Vec{f}, \widehat{\Vec{f}}),
\end{gather}
where Encoder and Decoder are realized by \acrshort{dnn}s and $L_{\text{rec}}$ denotes a feature reconstruction loss function such as the \acrshort{mse}. Within this paradigm, \acrshort{ssl} could impose an auxiliary task to the encoded features, the decoded features, or both. A generic representation of this process can be described by
\begin{gather}
    L_{\text{latent}}(\Vec{e}) = d(\Vec{e}, \text{Transform}(\Vec{e})) \label{eq:ssl_latent_loss}\\
    L_{\text{feature}}(\Vec{f}, \widehat{\Vec{f}}) = d(\text{Transform}(\widehat{\Vec{f}}), \text{Transform}(\Vec{f})), \label{eq:ssl_feature_loss}
\end{gather}
where each of $L_{\text{latent}}$ and $L_{\text{feature}}$ denotes the loss function of an auxiliary task. ``Transform'' refers to manipulation that provides distinctive goals to the auxiliary task. In context encoders \cite{pathak2016context}, for example, partial occlusion is applied to the input image, forcing the encoder to learn features that would extrapolate the occluded pixel values.

It should be noted that the labels used in the pretext tasks are readily available in the original dataset and therefore the training targets in \eref{eq:ssl_latent_loss} and \eref{eq:ssl_feature_loss} shall not incur additional labeling effort. More specifically, we shall design the task in such a way that it does not require the clean reference speech for real-world degraded noisy speech. This task shall ultimately enable an unsupervised loss function,
\begin{equation}
    L_{u}(\widehat{\Vec{S}}) = d(\widehat{\Vec{S}}, \text{Transform}(\widehat{\Vec{S}})).
\end{equation}
As opposed to \eref{eq:supervised_se}, this loss function can be used to train an \acrshort{se} system on real-world degraded speech data. In the next section, we will describe a procedure that enables this process.

\section{Method}
\begin{figure*}
    \centering
    \includegraphics[width=\linewidth, height=3in]{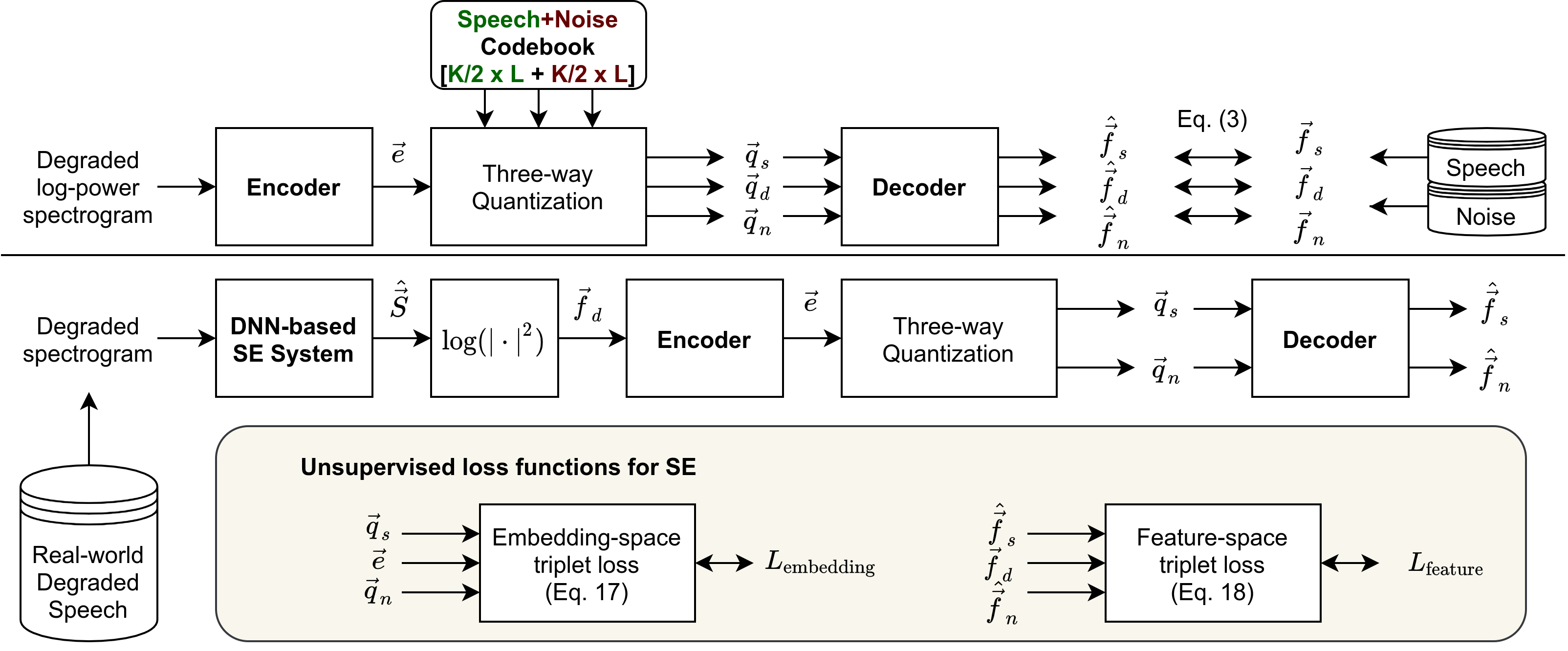}
    \caption{Supervised training procedure of the modified \acrshort{vqvae} using \emph{paired} data (\emph{top}) and unsupervised training procedure of a speech enhancement system using \emph{unpaired} data (\emph{bottom}).}
    \label{fig:ssl}
\end{figure*}
Our approach consists of two training stages. The first stage consists of training a modified \acrshort{vqvae} that is constrained to separate speech and noise from the degraded speech signal in both latent and feature domains. Then, we describe two loss functions derived from this model that can be used to train any \acrshort{dnn}-based \acrshort{se} systems in a semi-supervised manner. The unsupervised loss functions enable incorporation of real-world noisy speech during training. 

\subsection{Modified \acrshort{vqvae} for source separation}
Compared to traditional autoencoders, the \acrshort{vqvae} has an additional quantization step in the latent space that avoids issues like high variance \cite{oord2017neural}. The whole process can be described by
\begin{gather}
    \Vec{e} = \text{Encoder}(\Vec{f})\label{eq:vqvae_enc}\\
    \Vec{q} = \text{VQ}(\Vec{e}; \{\Vec{c}_i\}) = \argmin_{\Vec{c}_i}d(\Vec{e}, \Vec{c}_i)  \label{eq:vqvae_vq}\\
    \widehat{\Vec{f}} = \text{Decoder}(\Vec{q}), \label{eq:vqvae_dec}
\end{gather}
where $\{c_i\}, 1 \leq i \leq K$ is a set of $K$ learnable vectors, and $d(\cdot)$ is a distance function. We define $\Vec{f}$ to be the log-power spectra of degraded speech in vector form.

We design the \acrshort{vqvae} to do an acoustical source separation task with a codebook-lookup constraint. To achieve this goal, we partition the codebook $\{c_i\}$ in \eref{eq:vqvae_vq} into two equal halves,
\begin{gather}
    C_s = \{\Vec{c}_i\}, 1 \leq i \leq \frac{K}{2}\\
    C_n = \{\Vec{c}_i\}, \frac{K}{2} < i \leq K\\
    C_d = C_s \cup C_n = \{\Vec{c}_i\}, 1 \leq i \leq K.
\end{gather}
The quantization and decoding processes in \eref{eq:vqvae_vq} and \eref{eq:vqvae_dec} are then modified to produce three outputs,
\begin{gather}
    \Vec{q}_k = \text{VQ}(\Vec{e}; C_k) = \argmin_{C_k(i)}d(\Vec{e}, C_k(i)) \label{eq:mod_vqvae_vq} \\
    \widehat{\Vec{f}_k} = \text{Decoder}(\Vec{q}_k)\label{eq:mod_vqvae_dec} 
\end{gather}
where $k \in \{s, n, d\}$ denotes one of speech, noise, and degraded speech. 

\subsection{Encoder and decoder architectures}
\begin{figure}
    \centering
    \includegraphics[width=\linewidth]{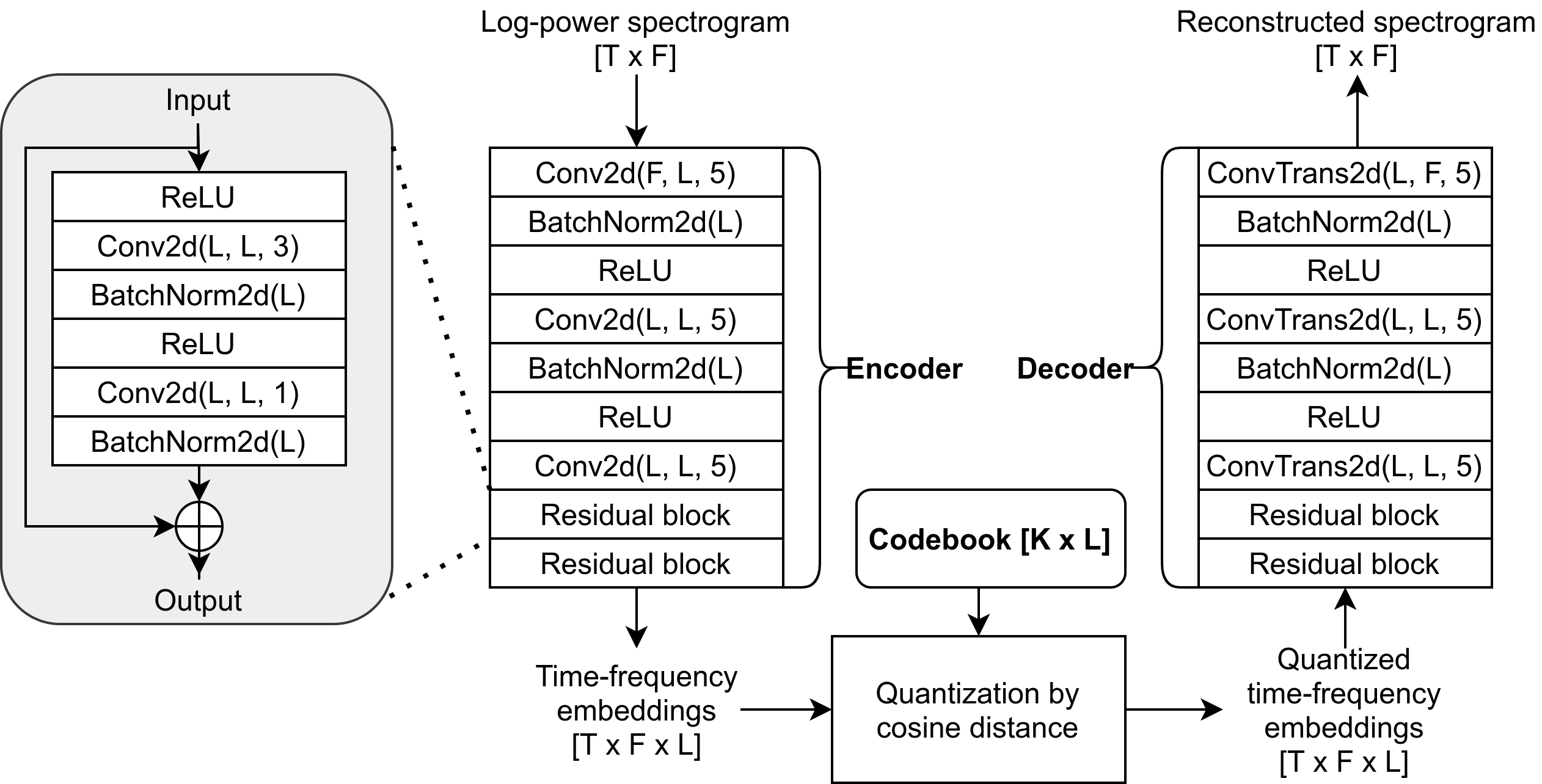}
    \caption{Flow diagram of our \acrshort{vqvae} system. The order of parameters follows the PyTorch convention.}
    \label{fig:vqvae}
\end{figure}
Both the encoder and decoder of our model are implemented using \acrfull{cnn}. Specifically, the encoder consists of three two-dimensional convolutional layers, each followed by two-dimensional batch normalization and \acrfull{relu}. The final convolutional layer is followed by two residual blocks; a residual block is defined as two convolutional layers with an additive connection between the input and the output of the final layer \cite{he2016deep}. The decoder's architecture is a mirror image of that of the encoder, with each convolutional layer replaced by a transposed convolutional layer.

The overall architecture is illustrated in \fref{fig:vqvae}. For a $T$-by-$F$ log-power spectrogram, each energy bin is transformed to a $L$-dimensional embedding by the encoder. The quantizer described in the previous step then transforms each embedding to its closest cluster center in terms of cosine distance. Finally, the decoder transforms quantized embeddings back to the log-power spectrogram.

\subsection{Training procedure for modified \acrshort{vqvae}}
The training procedure of the modified \acrshort{vqvae} is depicted in the top half of \fref{fig:ssl}. Log-power spectrograms of degraded speech signals pass through the \acrshort{vqvae} in order to obtain the embedding and reconstructed feature for each source. We train the \acrshort{vqvae} using the reconstruction loss in \eref{eq:supervised_se}, the VQ loss, and the commitment loss as described in \cite{oord2017neural}. Note that each loss now consists of three components based on the degraded speech, clean speech, and noise, respectively. Although the \acrfull{mse} was originally used in \cite{oord2017neural} as the distance function $d(\cdot)$ in \eref{eq:vqvae_vq}, we found that cosine distance made training more stable for this particular task. The VQ loss and the commitment loss are modified accordingly.


\subsection{Unsupervised loss functions for enhancement}
After training using \emph{paired} data and supervised loss functions, the \acrshort{vqvae} is frozen while training a speech enhancement system. Specifically, a supervised \acrshort{se} system takes in a degraded speech signal and outputs $\widehat{\Vec{S}}$, the \acrshort{stftm} of enhanced speech. It is then transformed to a log-power spectrogram and passed through the frozen \acrshort{vqvae} that outputs the continuous embedding, the quantized embeddings, and the decoded features in the process. We define the unsupervised embedding-space loss as
\begin{gather}\label{eq:embedding_triplet}
    L_{\text{embedding}}(\Vec{e}; \Theta) = max(d(\Vec{e}, \Vec{q}_s) - d(\Vec{e}, \Vec{q}_n) + m, 0),
\end{gather}
where $m$ is a constant and $d(\cdot)$ is the cosine distance. Note that the continuous embedding $\Vec{e}$ is used instead of the quantized embedding $\Vec{q}_d$ as the latter is not differentiable. Similarly, the unsupervised feature-space loss can be derived from the decoded features,
\begin{gather}\label{eq:feature_triplet}
    L_{\text{feature}}(\Vec{f}_d; \Theta) = max(d(\Vec{f}_d, \widehat{\Vec{f}}_s) - d(\Vec{f}_d, \widehat{\Vec{f}}_n) + m, 0),
\end{gather}
where $\widehat{\Vec{f}}_s$ and $\widehat{\Vec{f}}_n$ are decoded from their corresponding quantized embeddings using \eref{eq:mod_vqvae_dec}. Note that both losses are calculated per time-frequency bin. We believe that the source separation task imposed on the \acrshort{vqvae} makes $\Vec{q_s}$ and $\widehat{\Vec{f}}_s$ pseudo-positive targets, and $\Vec{q_n}$ and $\widehat{\Vec{f}}_n$ pseudo-negative targets. We used the triplet margin \cite{schroff2015facenet} because neither target is ideal.

The bottom half of \fref{fig:ssl} shows how to train a \acrshort{dnn}-based \acrshort{se} system using real-world data. After obtaining the enhanced log-power spectrogram from the system, the frozen \acrshort{vqvae} is used to calculate the continuous embedding, the quantized embeddings, and the reconstructed features. The unsupervised embedding loss can be calculated by \eref{eq:embedding_triplet}; the unsupervised feature loss can be calculated by \eref{eq:feature_triplet}. This loss is backpropagated to adapt the parameters of the \acrshort{se} system. If this system is also trained on paired data, the entire procedure is a semi-supervised training process.

In the next section, we will describe the experimental setup used to evaluate the effectiveness of these unsupervised losses for \acrlong{se}.


\section{Experimental setup}
\subsection{Dataset}
We used the clean speech of the Interspeech 2020 Deep Noise Suppression (DNS) Challenge dataset \cite{reddy2020interspeech} and the ESC-50 dataset \cite{piczak2015esc} for simulating \emph{paired data} in all our experiments. The DNS training set contains a total of 500 hours of clean speech. The ESC-50 dataset contains 50 different types of environmental sounds (noises). In our experiments, we used fractions of these datasets to synthesize the \emph{paired} data for training both the \acrshort{vqvae} and the supervised part of the \acrshort{se} system. The \emph{real-world noisy speech} or the \emph{unpaired data} was obtained from the Audioset dataset \cite{gemmeke2017audio}. Audio recordings in Audioset tagged with ``speech" class were further filtered by a sound event detector \cite{kumar2020sequential} to ensure that a large part of the recording contains speech along with other sounds.  All audio recordings are sampled at 16k Hz. The average SNR of the filtered Audioset data estimated by the WADA algorithm \cite{kim2008robust} is around 10 dB.

\subsection{Training procedure for VQ-VAE}
To train our \acrshort{vqvae}, we randomly sampled 1-second speech segment from the DNS dataset and 1-second noise segment from the ESC-50 dataset, respectively. We then mixed the two signals at a SNR randomly sampled from the range $[-10, 30)$ dB. The mixed signals are scaled to provide a dynamic range of 40 dB. The resulting degraded signal was the input to the \acrshort{vqvae}.

\subsection{Training and evaluation procedure for enhancement}
We used stacked Gated Recurrent Units described in \cite{xia2020weighted} as the baseline system for real-time speech enhancement in our experiments. Similar to the training procedure for the \acrshort{vqvae}, we simulated degraded speech from speech signals in the DNS dataset and noise from the ESC-50 dataset. The degraded-clean \emph{pairs} were used to train the \gls{se} system with the supervised loss function in \eref{eq:supervised_se}. 
We consider three different conditions for training the enhancement system: (1) \emph{Baseline}: the enhancement model is trained using only the paired data with supervised losses, (2) \emph{Paired-Unsupervised}: the unsupervised loss functions (either~\eref{eq:embedding_triplet} or~\eref{eq:feature_triplet}) are calculated from the \emph{paired} data, and (3) \emph{Unpaired-Unsupervised}: the unsupervised losses calculated from the real-world \emph{unpaired} data in addition to the \emph{paired} data. We summarize the setup of these systems in Table \ref{tab:systems}.
\begin{table}[th]
  \caption{System configurations}
  \label{tab:systems}
  \centering
  \begin{tabular}{c c c c}
    \toprule
    \multicolumn{1}{c}{\textbf{Method}} & 
    \multicolumn{1}{c}{\textbf{Training Data}} & \thead{\textbf{Unsupervised}\\\textbf{Loss Function}}\\
    \midrule
    Baseline & \emph{paired} & -\\
    Paired-Embedding & \emph{paired} & \eref{eq:embedding_triplet}\\
    Paired-Feature & \emph{paired} & \eref{eq:feature_triplet}\\
    Unpaired-Embedding & \emph{paired} \& \emph{unpaired} & \eref{eq:embedding_triplet}\\
    Unpaired-Feature & \emph{paired} \& \emph{unpaired} & \eref{eq:feature_triplet}\\
    \bottomrule
  \end{tabular}
\end{table}

To evaluate the quality of enhanced speech signals, we used the perceptual evaluation of speech quality (PESQ) \cite{rix2001perceptual} and scale-invariant signal-to-distortion ratio (SI-SDR) \cite{le2019sdr} metrics.

\begin{figure}
    \centering
    \includegraphics[width=\linewidth]{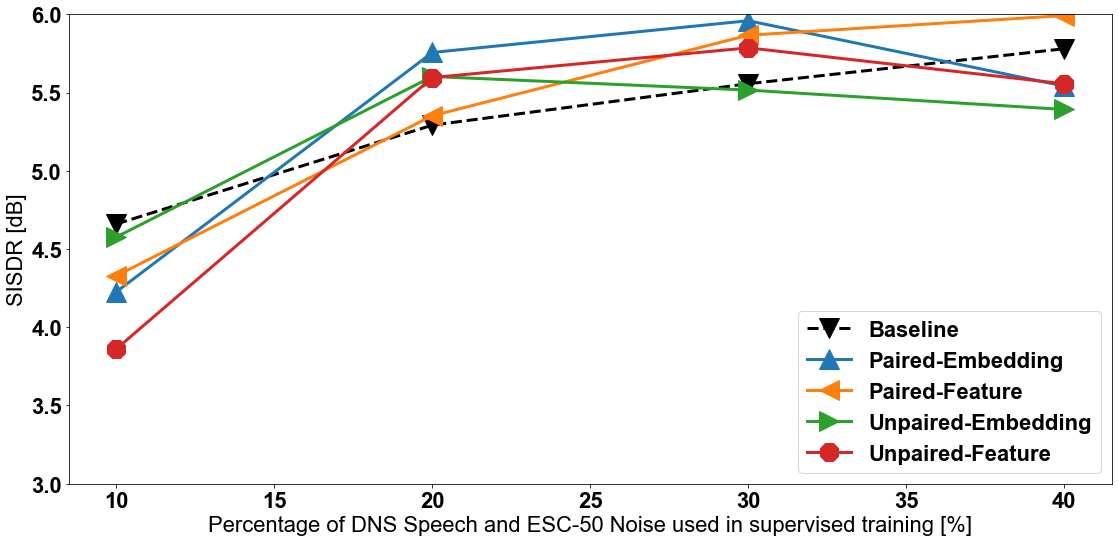}
    \includegraphics[width=\linewidth]{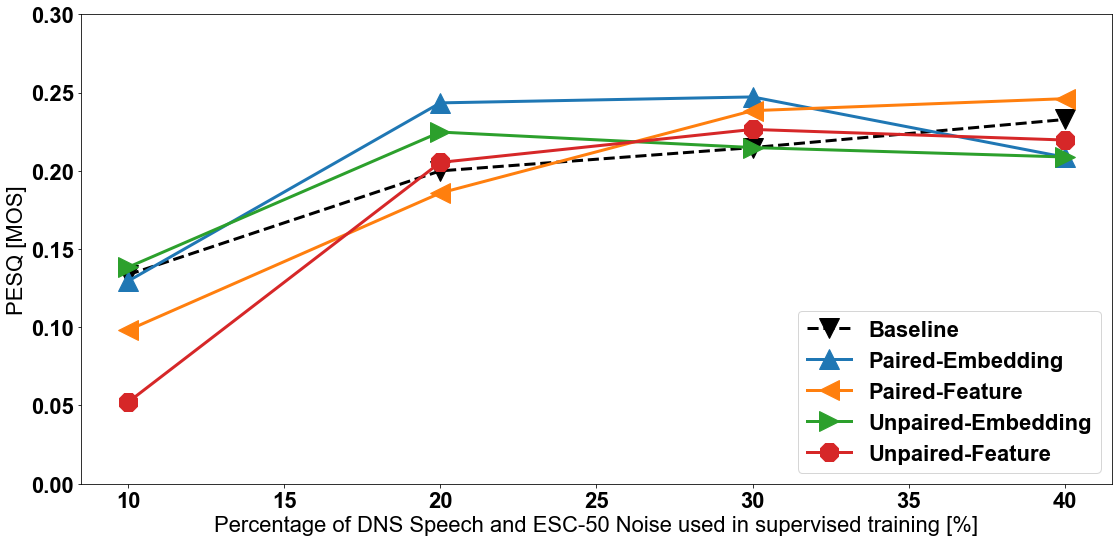}
    \caption{Absolute \gls{sisdr} improvement (\emph{top}) and \gls{pesq} improvement (\emph{bottom}) averaged across all evaluation conditions for \emph{seen} noise types during training. The averaged \gls{snr} and \gls{pesq} of unprocessed speech are 0 dB and 1.39 MOS, respectively.}
    \label{fig:data_fraction_seen}
\end{figure}
\begin{figure}
    \centering
    \includegraphics[width=\linewidth]{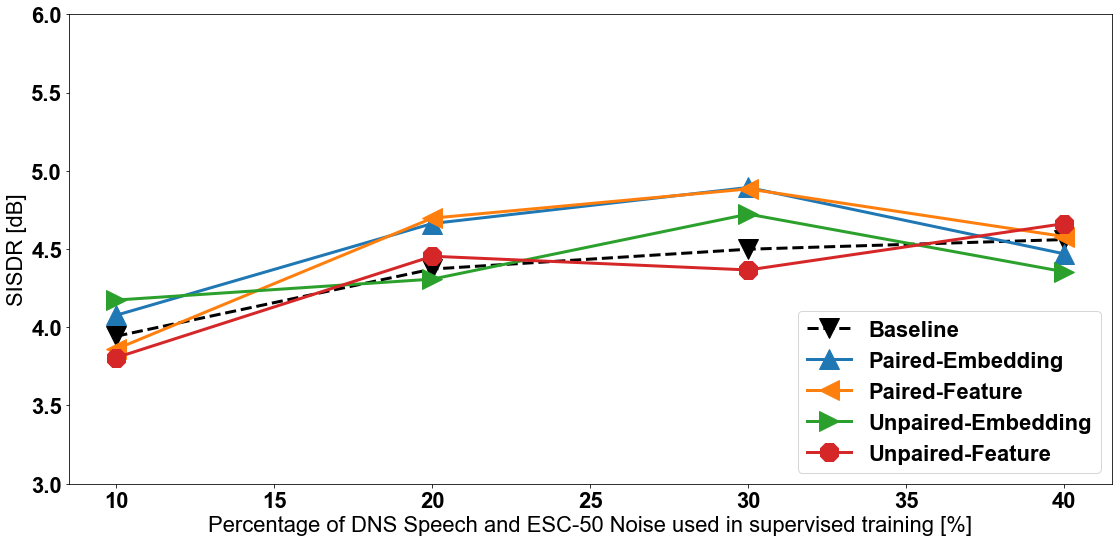}
    \includegraphics[width=\linewidth]{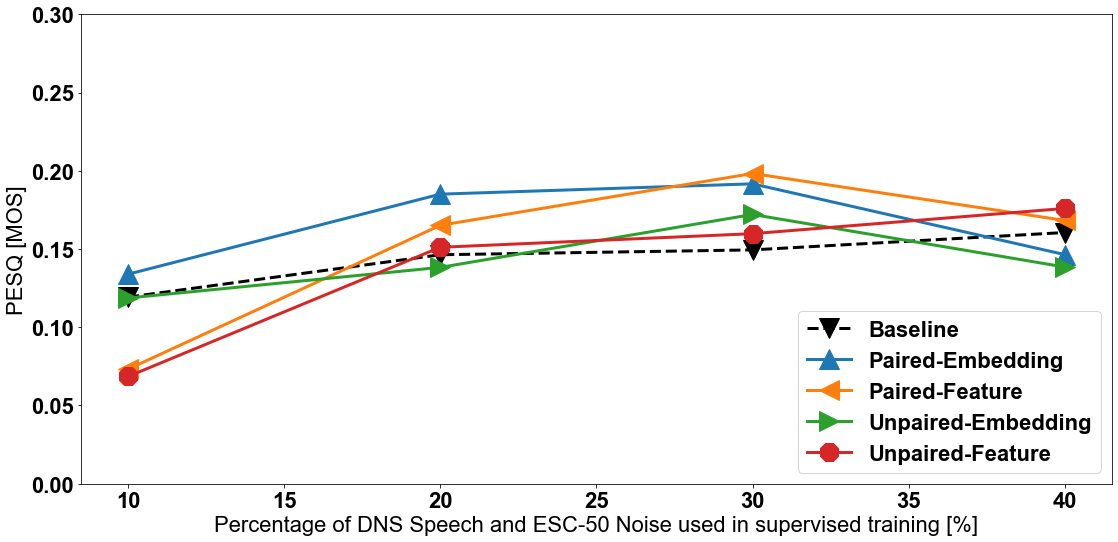}
    \caption{Absolute \gls{sisdr} improvement (\emph{top}) and \gls{pesq} improvement (\emph{bottom}) averaged across all evaluation conditions for \emph{unseen} noise types during training. The averaged \gls{snr} and \gls{pesq} of unprocessed speech are 0 dB and 1.41 MOS, respectively.}
    \label{fig:data_fraction_unseen}
\end{figure}

\section{Experimental results and discussions}
\begin{table*}[ht]\label{tab:20data}
\caption{Evaluation of speech enhancement systems trained on 20\% supervised data: \emph{seen} (\emph{unseen}) noise conditions}
\begin{tabular}{|c|cc|cc|cc|cc|cc|}
\hline
\multirow{3}{*}{Method} &
  \multicolumn{10}{c|}{SNR} \\
  \cline{2-11}
 &
  \multicolumn{2}{c|}{-10 dB} &
  \multicolumn{2}{c|}{-5 dB} &
  \multicolumn{2}{c|}{0 dB} &
  \multicolumn{2}{c|}{5 dB} &
  \multicolumn{2}{c|}{10 dB} \\
  \cline{2-11}
 &
  \multicolumn{1}{l|}{PESQ} &
  \multicolumn{1}{l|}{SI-SDR} &
  \multicolumn{1}{l|}{PESQ} &
  \multicolumn{1}{l|}{SI-SDR} &
  \multicolumn{1}{l|}{PESQ} &
  \multicolumn{1}{l|}{SI-SDR} &
  \multicolumn{1}{l|}{PESQ} &
  \multicolumn{1}{l|}{SI-SDR} &
  \multicolumn{1}{l|}{PESQ} &
  \multicolumn{1}{l|}{SI-SDR} \\ \hline
Degraded &
  \begin{tabular}[c]{@{}c@{}}1.16\\ (1.09)\end{tabular} &
  \multicolumn{1}{c|}{\begin{tabular}[c]{@{}c@{}}-9.98\\ (-10.0)\end{tabular}} &
  \begin{tabular}[c]{@{}c@{}}1.17\\ (1.15)\end{tabular} &
  \multicolumn{1}{c|}{\begin{tabular}[c]{@{}c@{}}-4.99\\ (-5.02)\end{tabular}} &
  \begin{tabular}[c]{@{}c@{}}1.31\\ (1.30)\end{tabular} &
  \multicolumn{1}{c|}{\begin{tabular}[c]{@{}c@{}}0.01\\ (-0.01)\end{tabular}} &
  \begin{tabular}[c]{@{}c@{}}1.53\\ (1.56)\end{tabular} &
  \multicolumn{1}{c|}{\begin{tabular}[c]{@{}c@{}}5.00\\ (4.99)\end{tabular}} &
  \begin{tabular}[c]{@{}c@{}}1.80\\ (1.95)\end{tabular} &
  \begin{tabular}[c]{@{}c@{}}10.0\\ (10.0)\end{tabular} \\ \hline
Baseline &
  \begin{tabular}[c]{@{}c@{}}1.29\\ (1.22)\end{tabular} &
  \multicolumn{1}{c|}{\begin{tabular}[c]{@{}c@{}}-3.13\\ (-2.85)\end{tabular}} &
  \begin{tabular}[c]{@{}c@{}}1.41\\ (1.27)\end{tabular} &
  \multicolumn{1}{c|}{\begin{tabular}[c]{@{}c@{}}1.46\\  (0.576)\end{tabular}} &
  \begin{tabular}[c]{@{}c@{}}1.55\\ (1.44)\end{tabular} &
  \multicolumn{1}{c|}{\begin{tabular}[c]{@{}c@{}}5.77\\ (4.25)\end{tabular}} &
  \begin{tabular}[c]{@{}c@{}}1.73\\ (1.74)\end{tabular} &
  \multicolumn{1}{c|}{\begin{tabular}[c]{@{}c@{}}9.61\\ (8.15)\end{tabular}} &
  \begin{tabular}[c]{@{}c@{}}1.98\\ (2.11)\end{tabular} &
  \begin{tabular}[c]{@{}c@{}}12.8\\ (11.6)\end{tabular} \\ \hline
Paired-Embedding &
  \textbf{\begin{tabular}[c]{@{}c@{}}1.31\\ (1.24)\end{tabular}} &
  \multicolumn{1}{c|}{\begin{tabular}[c]{@{}c@{}}\textbf{-2.44}\\ (-2.44)\end{tabular}} &
  \textbf{\begin{tabular}[c]{@{}c@{}}1.45\\ (1.32)\end{tabular}} &
  \multicolumn{1}{c|}{\textbf{\begin{tabular}[c]{@{}c@{}}2.16\\  (1.12)\end{tabular}}} &
  \textbf{\begin{tabular}[c]{@{}c@{}}1.60\\ (1.48)\end{tabular}} &
  \multicolumn{1}{c|}{\begin{tabular}[c]{@{}c@{}}\textbf{6.32}\\ (4.61)\end{tabular}} &
  \textbf{\begin{tabular}[c]{@{}c@{}}1.78\\ (1.78)\end{tabular}} &
  \multicolumn{1}{c|}{\begin{tabular}[c]{@{}c@{}}\textbf{9.93}\\ (8.28)\end{tabular}} &
  \textbf{\begin{tabular}[c]{@{}c@{}}2.05\\ (2.16)\end{tabular}} &
  \begin{tabular}[c]{@{}c@{}}12.9\\ (11.7)\end{tabular} \\ \hline
Paired-Feature &
  \begin{tabular}[c]{@{}c@{}}1.28\\ \textbf{(1.24)}\end{tabular} &
  \multicolumn{1}{c|}{\begin{tabular}[c]{@{}c@{}}-2.69\\ \textbf{(-2.33)}\end{tabular}} &
  \begin{tabular}[c]{@{}c@{}}1.41\\ (1.29)\end{tabular} &
  \multicolumn{1}{c|}{\begin{tabular}[c]{@{}c@{}}1.66\\  (1.08)\end{tabular}} &
  \begin{tabular}[c]{@{}c@{}}1.53\\ (1.47)\end{tabular} &
  \multicolumn{1}{c|}{\begin{tabular}[c]{@{}c@{}}5.75\\ \textbf{(4.65)}\end{tabular}} &
  \begin{tabular}[c]{@{}c@{}}1.71\\ (1.76)\end{tabular} &
  \multicolumn{1}{c|}{\begin{tabular}[c]{@{}c@{}}9.51\\ \textbf{(8.32)}\end{tabular}} &
  \begin{tabular}[c]{@{}c@{}}1.96\\ (2.11)\end{tabular} &
  \begin{tabular}[c]{@{}c@{}}12.6\\ (11.7)\end{tabular} \\ \hline
Unpaired-Embedding &
  \begin{tabular}[c]{@{}c@{}}1.30\\ (1.20)\end{tabular} &
  \multicolumn{1}{c|}{\begin{tabular}[c]{@{}c@{}}-2.64\\ (-2.84)\end{tabular}} &
  \begin{tabular}[c]{@{}c@{}}1.43 \\ (1.27)\end{tabular} &
  \multicolumn{1}{c|}{\begin{tabular}[c]{@{}c@{}}1.90\\  (0.530)\end{tabular}} &
  \begin{tabular}[c]{@{}c@{}}1.57\\ (1.43)\end{tabular} &
  \multicolumn{1}{c|}{\begin{tabular}[c]{@{}c@{}}6.05\\ (4.19)\end{tabular}} &
  \begin{tabular}[c]{@{}c@{}}1.76\\ (1.72)\end{tabular} &
  \multicolumn{1}{c|}{\begin{tabular}[c]{@{}c@{}}9.82\\ (7.98)\end{tabular}} &
  \begin{tabular}[c]{@{}c@{}}2.02\\ (2.12)\end{tabular} &
  \begin{tabular}[c]{@{}c@{}}12.9\\ (11.6)\end{tabular} \\ \hline
Unpaired-Feature &
  \begin{tabular}[c]{@{}c@{}}\textbf{1.31}\\ (1.21)\end{tabular} &
  \multicolumn{1}{c|}{\begin{tabular}[c]{@{}c@{}}-2.56\\ (-2.88)\end{tabular}} &
  \begin{tabular}[c]{@{}c@{}}1.43 \\ (1.27)\end{tabular} &
  \multicolumn{1}{c|}{\begin{tabular}[c]{@{}c@{}}1.83\\ (0.543)\end{tabular}} &
  \begin{tabular}[c]{@{}c@{}}1.56\\ (1.45)\end{tabular} &
  \multicolumn{1}{c|}{\begin{tabular}[c]{@{}c@{}}6.05\\ (4.37)\end{tabular}} &
  \begin{tabular}[c]{@{}c@{}}1.73\\ (1.75)\end{tabular} &
  \multicolumn{1}{c|}{\begin{tabular}[c]{@{}c@{}}9.73\\ (8.25)\end{tabular}} &
  \begin{tabular}[c]{@{}c@{}}1.97\\ (2.13)\end{tabular} &
  \textbf{\begin{tabular}[c]{@{}c@{}}13.0\\ (11.9)\end{tabular}} \\ \hline
\end{tabular}
\end{table*}

\subsection{Effect of the amount of \emph{paired} data}
We present the absolute improvement of all \acrshort{se} systems under the \emph{seen} noise condition as a function of the amount of \emph{paired} training data in \fref{fig:data_fraction_seen}. With the minimum \emph{paired} data (10\%), the unsupervised losses based training were not able to improve over the supervised baseline; in fact, many performed noticeably worse than the baseline. As the amount of \emph{paired} data increased to 20\% of DNS speech and ESC-50 noise, all unsupervised loss functions were able to largely improve and surpassed the baseline performance. This indicates that a decent amount of \emph{paired} data is necessary for making the \acrshort{vqvae} learn a reliable representation of speech and noise. Finally, as more amount of \emph{paired} data was presented in training, the significance of unsupervised losses goes down. This suggests that the supervised loss function eventually outweighs the unsupervised losses.

\subsection{Generalization to unseen noise types}
We present the absolute improvement of all \acrshort{se} systems under the \emph{unseen} noise condition as a function of the amount of \emph{paired} training data in \fref{fig:data_fraction_unseen}. We note the similar trend as observed in the results for the \emph{seen} noise conditions: unsupervised loss functions require a certain amount of supervised training to benefit the system. The overall performance compared to the \emph{seen} noise condition is generally worse and improves slower as the amount of \emph{paired} data increased. This phenomenon is generally true for supervised \acrshort{se} systems. At 30\% of supervised data, however, we observe similar improvement to the \emph{seen} noise condition by including the unsupervised losses calculated from the \emph{paired} data. This indicates that the unsupervised losses calculated from the \emph{paired} data is generalizable to unseen noise conditions.

\subsection{Effect of unsupervised loss functions}
As \fref{fig:data_fraction_seen} and \fref{fig:data_fraction_unseen} revealed that 20\% of supervised data is the minimum from our setting that the \acrshort{se} systems start benefitting from unsupervised loss functions, we present the detailed evaluation across noise conditions in Table \ref{tab:20data}. Results show that the paired-embedding loss is the best across most SNR conditions. The paired-feature loss is slightly more superior under some unseen noise types. The unpaired loss functions had more impact at higher SNRs. We believe that this could be because the filtered Audioset has relatively high SNR.

\subsection{Learned embedding margin}
\begin{figure}
    \centering
    \includegraphics[width=\linewidth]{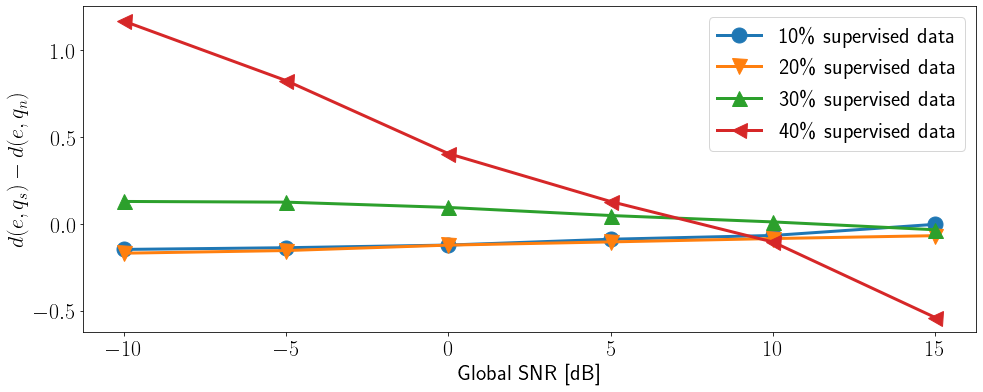}
    \caption{Averaged triplet margin on the validation set as a function of global SNR. Lines with more negative slopes correspond to better learned representation. The margin was calculated on the validation set in three-dimensional latent space. The standard deviations from smallest amount of data to largest amount of data were 0.55, 0.28, 0.31, and 1.53, respectively.}
    \label{fig:margin}
\end{figure}

To verify that the source separation task imposed on the \acrshort{vqvae} was effective, we present the averaged triplet margin calculated on the validation set in \fref{fig:margin}. As defined in \eref{eq:embedding_triplet}, the margin should be high when global SNR is low, and the margin should be low when global SNR is high. As \fref{fig:margin} shows, using 10\% and 20\% supervised data were not enough to learn the correct relationship. While using 30\% supervised data worked, using 40\% data made a more drastic improvement. This shows that the more training data the better the \acrshort{vqvae} learns the \acrshort{ssl} tasks, which in turn would improve the quality of unsupervised losses for \acrshort{se}.

\section{Conclusions}
In this paper, we introduced two novel unsupervised loss functions for speech enhancement that were enabled by a modified vector-quantized variational autoencoder and a self-supervised learning task. We showed that the loss functions calculated on supervised data were able to improve supervised speech enhancement systems when the amount of training data is small. We also showed that the loss functions calculated on real-world noisy speech data were able to improve the supervised \acrshort{se} systems in some noise conditions. In the future, we plan on fine-tuning the \acrshort{vqvae} on enhanced speech data. We will also explore sampling techniques of real-world data to better match the evaluation condition.
\newpage
\bibliographystyle{IEEEbib}
\bibliography{mybib}

\begin{thebibliography}{10}

\bibitem{wang2014training}
Yuxuan Wang, Arun Narayanan, and DeLiang Wang,
\newblock ``On training targets for supervised speech separation,''
\newblock {\em IEEE/ACM transactions on audio, speech, and language
  processing}, vol. 22, no. 12, pp. 1849--1858, 2014.

\bibitem{luo2019conv}
Y.~Luo and N.~Mesgarani,
\newblock ``{Conv-TasNet}: Surpassing ideal time--frequency magnitude masking
  for speech separation,''
\newblock {\em IEEE/ACM transactions on audio, speech, and language
  processing}, pp. 1256--1266, 2019.

\bibitem{wisdom2020unsupervised}
Scott Wisdom, Efthymios Tzinis, Hakan Erdogan, Ron~J. Weiss, Kevin Wilson, and
  John~R. Hershey,
\newblock ``Unsupervised sound separation using mixture invariant training,''
\newblock in {\em NeurIPS}, 2020.

\bibitem{fujimura2021noisy}
Takuya Fujimura, Yuma Koizumi, Kohei Yatabe, and Ryoichi Miyazaki,
\newblock ``Noisy-target training: A training strategy for dnn-based speech
  enhancement without clean speech,''
\newblock {\em arXiv preprint arXiv:2101.08625}, 2021.

\bibitem{9054204}
A.~A. {Catellier} and S.~D. {Voran},
\newblock ``Wawenets: A no-reference convolutional waveform-based approach to
  estimating narrowband and wideband speech quality,''
\newblock in {\em IEEE International Conference on Acoustics, Speech and Signal
  Processing (ICASSP)}, 2020, pp. 331--335.

\bibitem{martin2018deep}
Juan~Manuel Martin-Donas, Angel~Manuel Gomez, Jose~A Gonzalez, and Antonio~M
  Peinado,
\newblock ``A deep learning loss function based on the perceptual evaluation of
  the speech quality,''
\newblock {\em IEEE Signal processing letters}, vol. 25, no. 11, pp.
  1680--1684, 2018.

\bibitem{zhao2018perceptually}
Y.~Zhao, B.~Xu, R.~Giri, and T.~Zhang,
\newblock ``Perceptually guided speech enhancement using deep neural
  networks,''
\newblock in {\em IEEE International Conference on Acoustics, Speech and Signal
  Processing}, 2018, pp. 5074--5078.

\bibitem{wang2020self}
Yu-Che Wang, Shrikant Venkataramani, and Paris Smaragdis,
\newblock ``Self-supervised learning for speech enhancement,''
\newblock {\em arXiv preprint arXiv:2006.10388}, 2020.

\bibitem{Xiang2020}
Yang Xiang and Changchun Bao,
\newblock ``A parallel-data-free speech enhancement method using
  multi-objective learning cycle-consistent generative adversarial network,''
\newblock {\em IEEE/ACM Transactions on Audio, Speech, and Language
  Processing}, vol. 28, pp. 1826--1838, 2020.

\bibitem{zhu2017unpaired}
Jun-Yan Zhu, Taesung Park, Phillip Isola, and Alexei~A Efros,
\newblock ``Unpaired image-to-image translation using cycle-consistent
  adversarial networks,''
\newblock in {\em Proceedings of the IEEE ICCV}, 2017, pp. 2223--2232.

\bibitem{pathak2016context}
Deepak Pathak, Philipp Krahenbuhl, Jeff Donahue, Trevor Darrell, and Alexei~A
  Efros,
\newblock ``Context encoders: Feature learning by inpainting,''
\newblock in {\em Proceedings of the IEEE conference on computer vision and
  pattern recognition}, 2016, pp. 2536--2544.

\bibitem{oord2017neural}
Aaron van~den Oord, O~Vinyals, and K~Kavukcuoglu,
\newblock ``Neural discrete representation learning,''
\newblock in {\em 31st International Conference on Neural Information
  Processing Systems}, 2017, p. 6309–6318.

\bibitem{he2016deep}
Kaiming He, Xiangyu Zhang, Shaoqing Ren, and Jian Sun,
\newblock ``Deep residual learning for image recognition,''
\newblock in {\em Proceedings of the IEEE conference on computer vision and
  pattern recognition}, 2016, pp. 770--778.

\bibitem{schroff2015facenet}
F.~Schroff, D.~Kalenichenko, and J.~Philbin,
\newblock ``Facenet: A unified embedding for face recognition and clustering,''
\newblock in {\em Proceedings of the IEEE conference on computer vision and
  pattern recognition}, 2015, pp. 815--823.

\bibitem{reddy2020interspeech}
Chandan~KA Reddy, E~Beyrami, H~Dubey, V~Gopal, R~Cheng, R~Cutler, S~Matusevych,
  R~Aichner, A~Aazami, S~Braun, et~al.,
\newblock ``{The INTERSPEECH 2020 Deep Noise Suppression Challenge: Datasets,
  Subjective Speech Quality and Testing Framework},''
\newblock {\em arXiv preprint arXiv:2001.08662}, 2020.

\bibitem{piczak2015esc}
Karol~J Piczak,
\newblock ``{ESC: Dataset for environmental sound classification},''
\newblock in {\em Proceedings of the 23rd ACM international conference on
  Multimedia}, 2015, pp. 1015--1018.

\bibitem{gemmeke2017audio}
J.~Gemmeke, D.~Ellis, D.~Freedman, A.~Jansen, W.~Lawrence, R~Channing Moore,
  M.~Plakal, and M.~Ritter,
\newblock ``Audio {S}et: An ontology and human-labeled dataset for audio
  events,''
\newblock in {\em ICASSP}, 2017.

\bibitem{kumar2020sequential}
Anurag Kumar and Vamsi Ithapu,
\newblock ``A sequential self teaching approach for improving generalization in
  sound event recognition,''
\newblock in {\em International Conference on Machine Learning}. PMLR, 2020,
  pp. 5447--5457.

\bibitem{kim2008robust}
C.~Kim and R~M Stern,
\newblock ``Robust signal-to-noise ratio estimation based on waveform amplitude
  distribution analysis,''
\newblock in {\em Ninth Annual Conference of the International Speech
  Communication Association}, 2008.

\bibitem{xia2020weighted}
Yangyang Xia, Sebastian Braun, Chandan~KA Reddy, Harishchandra Dubey, Ross
  Cutler, and Ivan Tashev,
\newblock ``Weighted speech distortion losses for neural-network-based
  real-time speech enhancement,''
\newblock in {\em IEEE International Conference on Acoustics, Speech and Signal
  Processing (ICASSP)}. IEEE, 2020, pp. 871--875.

\bibitem{rix2001perceptual}
A~W Rix, J~G Beerends, M~P Hollier, and A~P Hekstra,
\newblock ``{Perceptual evaluation of speech quality (PESQ)-a new method for
  speech quality assessment of telephone networks and codecs},''
\newblock in {\em IEEE ICASSP}, 2001, vol.~2, pp. 749--752.

\bibitem{le2019sdr}
Jonathan Le~Roux, Scott Wisdom, Hakan Erdogan, and John~R Hershey,
\newblock ``{SDR--half-baked or well done?},''
\newblock in {\em IEEE International Conference on Acoustics, Speech and Signal
  Processing (ICASSP)}, 2019, pp. 626--630.

\end{thebibliography}

\end{document}